\documentclass[12pt]{article}
\begin{document}
\begin{center}
{\large\bf The Accelerating Universe: A Gravitational Explanation}
\vskip 0.3 true in
{\large J. W. Moffat}
\vskip 0.3 true in
{\it Department of Physics, University of Toronto,
Toronto, Ontario M5S 1A7, Canada}
\end{center}
\begin{abstract}%
The problem of explaining the acceleration of the expansion of the
universe and the observational and theoretical difficulties associated with
dark matter and dark energy are discussed. The possibility that GR does not
correctly describe the large-scale structure of the universe is considered
and an alternative gravity theory is proposed as a possible resolution to
the problems.

\end{abstract}
\vskip 0.2 true in
e-mail: moffat@medb.physics.utoronto.ca


\section{Introduction}

The recent surprising observational discovery that the expansion of
the universe is accelerating~\cite{Perlmutter} has led to an increasing
theoretical effort to understand this phenomenon. The attempt to
interpret the data by postulating a non-zero positive cosmological constant
is not satisfactory, because it is confronted by the
two serious issues of why the estimates from the standard model and
quantum field theory lead to preposterously large values of the
cosmological constant, and the coincidence of matter and dark
energy dominance today~\cite{Caldwell}.

If we simply postulate a repulsive force in the universe associated with a
charge density, then we might expect that this force could be responsible
for generating the acceleration of the universe. However, for a
homogeneous and isotropic universe the net charge density would be
zero, although for a finite range force with a small mass
there will exist a non-zero charge density~\cite{Woodard}. The effect of a
Maxwell-type force would be to lower or raise the total energy, leaving the
form of the Friedmann equation unchanged. Thus, we would still have to
invoke exotic forms of energy with an equation of state, $p=w\rho$, where
$w$ is negative and violates the positive energy theorems. For a
non-zero cosmological constant $w=-1$.

In addition to the dark energy problem, we are still confronted with the
puzzle of dark matter. Any observational detection of a dark matter
candidate has eluded us and the fits to galaxy halos using dark matter
models are based on several parameters depending on the size of the
galaxy being fitted. The dark matter model predictions disagree with
observable properties of galaxies~\cite{Sellwood}.

Another problem with the dark energy hypothesis is the serious challenge
to present day particle physics and string theory from the existence of
cosmological horizons, which arise in an eternally accelerating
universe~\cite{Fischler}. Several resolutions of this problem have been
proposed~\cite{Cline} but a cosmological horizon does produce a potentially
serious crisis for modern particle physics and string theory.

Challenging experimental results are often the precursors of
a shifting of scientific paradigms. We must now entertain the
prospect that the discovery of the mechanism driving the acceleration of
the universe can profoundly change our description of the universe.

Given the uneasy tension existing between observational evidence for the
acceleration of the universe and the mystery of what
constitutes dark matter and dark energy, we are tempted to reconsider the
question of whether Einstein's gravity theory (GR) is correct for the large
scale structure of the universe. It agrees well with local solar
system experimental tests and for the data obtained for observations of the
binary pulsar PSR 1913+16. However, this does not preclude the possibility
of a breakdown of the conventional Einstein equations for the
large-scale structure of the universe. The standard GR cosmological model
agrees well with the abundances of light elements from big bang
nucleosynthesis (BBN), and the evolution of the spectrum of primordial
density fluctuations, yielding the observed spectrum of temperature
anisotropies in the cosmic microwave background (CMB). Also, the age of the
universe and the power spectrum of large-scale structure agree reasonably
well with the standard cosmological model. However, it could be that
additional repulsive gravitational effects from an alternative gravity
theory could agree with all of the results in the early universe and yet
lead to significant effects in the present universe accounting for its
acceleration~\cite{Carroll}.

When contemplating alternative gravity theories, one is impressed with the
mathematical and physical robustness of GR. It is not easy to change the
structure of GR without running into consistency problems. A
fundamental change in the predictions of the observational data will
presumably only come about from a non-trivial alteration of the
mathematical and geometrical formalism that constitutes GR. From the
cosmological standpoint, such theories as
Jordan-Brans-Dicke~\cite{Weinberg} theories of gravity will not radically
change the Friedmann equation in the present universe. Recent developments
in brane-bulk cosmological models~\cite{Binetruy} have led to alterations
of the Friedmann equation but only for the very early universe
corresponding to high energies.

This tempts us to return to a
physically non-trivial extension of GR called the nonsymmetric gravity
theory (NGT). This theory was extensively studied over a period of
years, and a version of the theory
was discovered that was free of several possible inconsistencies such as
ghost poles, tachyons, exotic asymptotic behaviour and other
instabilities~\cite{Moffat,Moffat2,Moffat3,Moffat4}. Further research is
needed to fully understand such problems as its Cauchy development and the
deeper meaning of the basic gauge symmetries underlying the theory.

As we shall see in the following, NGT can describe the current data on
the accelerating universe and the dark matter halos of galaxies,
gravitational lensing and cluster behaviour, as well as the standard
results such as BBN, the solar system tests and the binary pulsar PSR
1913+16, without invoking the need for dominant, exotic dark matter and
dark energy.

\section{NGT Action and Field Equations}

The nonsymmetric $g_{\mu\nu}$ and
$\Gamma^\lambda_{\mu\nu}$ are defined
by~\cite{Moffat,Moffat2,Moffat3,Moffat4,Moffat5}:
\begin{equation}
g_{(\mu\nu)}={1\over 2}(g_{\mu\nu}+g_{\nu\mu}),\quad g_{[\mu\nu]}= {1\over
2}(g_{\mu\nu}-g_{\nu\mu}), \end{equation} and \begin{equation}
\Gamma^\lambda_{\mu\nu}=\Gamma^\lambda_{(\mu\nu)}
+\Gamma^\lambda_{[\mu\nu]}.
\end{equation}
The contravariant tensor
$g^{\mu\nu}$ is defined in terms of the equation
\begin{equation}
\label{inverse}
g^{\mu\nu}g_{\sigma\nu}=g^{\nu\mu}g_{\nu\sigma}={\delta^\mu}_\sigma.
\end{equation}

The Lagrangian density is given by
\begin{equation}
{\cal L}_{NGT}={\cal L}+{\cal L}_M,
\end{equation}
where
\begin{eqnarray}
\label{NGTLagrangian}
{\cal L}={\bf g}^{\mu\nu}R_{\mu\nu}(W)-2\Lambda\sqrt{-g}
-{1\over 4}\mu^2{\bf g}^{\mu\nu}g_{[\nu\mu]}\nonumber\\
-{1\over 6}{\bf
g}^{\mu\nu} W_\mu W_\nu+{\bf g}^{\mu\nu}J_{[\mu}\phi_{\nu]},
\end{eqnarray}
and ${\cal L}_M$ is the matter Lagrangian density ($G=c=1$):
\begin{equation}
{\cal L}_M=-8\pi g^{\mu\nu}{\bf T}_{\mu\nu}.
\end{equation}
Here, ${\bf g}^{\mu\nu}=\sqrt{-g}g^{\mu\nu}$,
$g=\hbox{Det}(g_{\mu\nu})$, $\Lambda$ is the cosmological constant and
$R_{\mu\nu}(W)$ is the NGT contracted curvature tensor:
\begin{equation}
R_{\mu\nu}(W)=W^\beta_{\mu\nu,\beta} - {1\over
2}(W^\beta_{\mu\beta,\nu}+W^\beta_{\nu\beta,\mu}) -
W^\beta_{\alpha\nu}W^\alpha_{\mu\beta} +
W^\beta_{\alpha\beta}W^\alpha_{\mu\nu},
\end{equation} defined in terms of
the unconstrained nonsymmetric connection
\begin{equation}
\label{Wequation}
W^\lambda_{\mu\nu}=\Gamma^\lambda_{\mu\nu}-{2\over
3}{\delta^\lambda}_\mu W_\nu,
\end{equation}
where
\[
W_\mu={1\over
2}(W^\lambda_{\mu\lambda}-W^\lambda_{\lambda\mu}). \]
Eq.(\ref{Wequation})
leads to the result
\[ \Gamma_\mu=\Gamma^\lambda_{[\mu\lambda]}=0.
\]
The
contracted tensor $R_{\mu\nu}(W)$ can be written as
\[
R_{\mu\nu}(W)=R_{\mu\nu}(\Gamma)+\frac{2}{3}W_{[\mu,\nu]}, \] where \[
R_{\mu\nu}(\Gamma ) = \Gamma^\beta_{\mu\nu,\beta} -{1\over 2}
\left(\Gamma^\beta_{(\mu\beta),\nu} + \Gamma^\beta_{(\nu\beta),\mu}\right)
- \Gamma^\beta_{\alpha\nu} \Gamma^\alpha_{\mu\beta} +
\Gamma^\beta_{(\alpha\beta)}\Gamma^\alpha_{\mu\nu}. \]

The term in Eq.(\ref{NGTLagrangian}):
\begin{equation}
\label{Lagrangemult}
{\bf g}^{\mu\nu}J_{[\mu}\phi_{\nu]},
\end{equation}
contains the Lagrange multiplier fields $\phi_\mu$ and a source vector
$J_\mu$.

A variation of the action
\[
S=\int d^4x{\cal L}_{\hbox{NGT}}
\]
yields the field equations in the presence of matter sources:
\begin{equation}
\label{Gequation}
G_{\mu\nu} (W)+\Lambda g_{\mu\nu}+S_{\mu\nu}
=8\pi (T_{\mu\nu}+K_{\mu\nu}),
\end{equation}
\begin{equation}
\label{divg}
{{\bf g}^{[\mu\nu]}}_{,\nu}=-\frac{1}{2}{\bf g}^{(\mu\alpha)}W_\alpha,
\end{equation}
\begin{equation}
{{\bf g}^{\mu\nu}}_{,\sigma}+{\bf g}^{\rho\nu}W^\mu_{\rho\sigma}
+{\bf g}^{\mu\rho}
W^\nu_{\sigma\rho}-{\bf g}^{\mu\nu}W^\rho_{\sigma\rho}
$$ $$
+{2\over 3}\delta^\nu_\sigma{\bf g}^{\mu\rho}W^\beta_{[\rho\beta]}
+{1\over 6}({\bf g}^{(\mu\beta)}W_\beta\delta^\nu_\sigma
-{\bf g}^{(\nu\beta)}W_\beta\delta^\mu_\sigma)=0.
\end{equation}
Here, we have $G_{\mu\nu}=R_{\mu\nu} - {1\over 2} g_{\mu\nu} R$, and
\begin{equation}
S_{\mu\nu}=\frac{1}{4}\mu^2(g_{[\mu\nu]}
+{1\over 2}g_{\mu\nu}g^{[\sigma\rho]}
g_{[\rho\sigma]}+g^{[\sigma\rho]}g_{\mu\sigma}g_{\rho\nu})-
\frac{1}{6}(W_\mu W_\nu-\frac{1}{2}g_{\mu\nu}g^{\alpha\beta}W_\alpha W_\beta).
\end{equation}
Moreover, the contribution from the variation of (\ref{Lagrangemult})
with respect to $g^{\mu\nu}$ and $\sqrt{-g}$ is given by
\begin{equation}
\label{Kequation}
K_{\mu\nu}=-\frac{1}{8\pi}[J_{[\mu}\phi_{\nu]}-\frac{1}{2}
g_{\mu\nu}(g^{[\alpha\beta]}J_{[\alpha}\phi_{\beta]})].
\end{equation}

The variation of $\phi_\mu$ yields the constraint equations
\begin{equation}
\label{gskewconstraint}
{\bf g}^{[\mu\nu]}J_\nu=0.
\end{equation}
We have not varied the source vector $J_\mu$.

If we use (\ref{gskewconstraint}), then (\ref{Kequation}) becomes
\begin{equation}
\label{K2}
K_{[\mu\nu]}=-\frac{1}{8\pi}J_{[\mu}\phi_{\nu]}.
\end{equation}

If we specify $J_\mu$ to be $J_\mu=(0,0,0,J_0)$, then
(\ref{gskewconstraint}) corresponds to the three constraint equations
\begin{equation}
\label{constraints2} {\bf g}^{[i0]}=0.
\end{equation}
 
After eliminating the Lagrange multiplier field $\phi_\mu$ from the field
equations (\ref{Gequation}), we get
\begin{equation}
 G_{(\mu\nu)}(W)+\Lambda
g_{(\mu\nu)}+S_{(\mu\nu)} =8\pi T_{(\mu\nu)},
\end{equation}
\begin{equation}
\epsilon^{\mu\nu\alpha\beta}J_\alpha(G_{[\mu\nu]}(W)+\Lambda
g_{[\mu\nu]}+S_{[\mu\nu]}) =8\pi\epsilon^{\mu\nu\alpha\beta}J_\alpha
T_{[\mu\nu]},
\end{equation}
where $\epsilon^{\mu\nu\alpha\beta}$ is the Levi-Civita symbol.

The generalized Bianchi identities
\begin{equation}
[{\bf g}^{\alpha\nu}G_{\rho\nu}(\Gamma)+{\bf g}^{\nu\alpha}
G_{\nu\rho}(\Gamma)]_{,\alpha}+{g^{\mu\nu}}_{,\rho}{\bf G}_{\mu\nu}=0,
\end{equation}
give rise to the matter response equations
\begin{equation}
g_{\mu\rho}{{\bf T}^{\mu\nu}}_{,\nu}+g_{\rho\mu}{{\bf T}^{\nu\mu}}_{,\nu}
+(g_{\mu\rho,\nu}+g_{\rho\nu,\mu}-g_{\mu\nu,\rho}){\bf T}^{\mu\nu}=0.
\end{equation}

A study of the linear approximation has proved that the present version of
NGT described above does not possess any ghost poles or tachyons in the
linear approximation~\cite{Moffat3}. This cures the inconsistencies
discovered by Damour, Deser and McCarthy in an earlier version of
NGT~\cite{Damour}. Moreover, the instability discovered by
Clayton~\cite{Clayton} for both massless and massive NGT in a Hamiltonian
formalism, associated with three of the six possible propagating degrees of
freedom in the skew symmetric sector, is eliminated from the theory.
This is implemented in the NGT action by the covariant constraint equations
(\ref{gskewconstraint}).

\section{Cosmological Solutions}

For the case of a spherically symmetric field, the canonical form of
$g_{\mu\nu}$ in NGT is given by
\begin{equation}
g_{\mu\nu}=\left(\matrix{-\alpha&0&0&w\cr
0&-\beta&f\hbox{sin}\theta&0\cr 0&-f\hbox{sin}\theta&
-\beta\hbox{sin}^2
\theta&0\cr-w&0&0&\gamma\cr}\right),
\end{equation}
where $\alpha,\beta,\gamma$ and $w$ are functions of $r$ and $t$.
We have
\[
\sqrt{-g}=\hbox{sin}\theta[(\alpha\gamma-w^2)(\beta^2+f^2)]^{1/2}.
\]
For a comoving coordinate system, we obtain for the velocity vector
$u^\mu$ which satisfies the normalization condition $g_{(\mu\nu)}u^\mu
u^\nu=1$:
\begin{equation}
\label{comovingvelocity}
u^0=\frac{1}{\sqrt{\gamma}},\quad u^r=u^{\theta}=u^{\phi}=0.
\end{equation}
From Eq.(\ref{constraints2}), we get $w=0$ and only the $g_{[23]}$
component of $g_{[\mu\nu]}$ is different from zero. The vector $W_\mu$ can
be determined from
\begin{equation}
\label{W2equation} W_\mu=-{2\over
\sqrt{-g}}s_{\mu\rho}{{\bf g}^{[\rho\sigma]}}_{,\sigma},
\end{equation}
where $s_{\mu\alpha}g^{(\alpha\nu)}=\delta^\nu_\mu$. For the spherically
symmetric field with $w=0$, it follows from (\ref{constraints2}) and
(\ref{W2equation}) that $W_\mu=0$.

The energy-momentum tensor for a fluid is
\begin{equation}
\label{eq:energytensor}
T^{\mu\nu}=(\rho+p)u^\mu u^\nu - pg^{\mu\nu}+B^{[\mu\nu]},
\end{equation}
where $B^{[\mu\nu]}$ is a skew symmetric source tensor.

We can prove from a Killing vector analysis that for a homogeneous and
isotropic universe massless NGT requires that
$f(r,t)=0$~\cite{Moffat5}. It follows that for massless NGT all
strictly homogeneous and isotropic solutions of NGT cosmology reduce to the
Friedmann-Robertson-Walker (FRW) solutions of GR. For the case of massive
NGT, it will be possible to obtain strictly homogeneous and isotropic
solutions. In the following, we shall solve the field equations for a
spherically symmetric inhomogenous universe and then approximate the
solution by assuming that the inhomogeneities are small.
We expand the metric $g_{(\mu\nu)}$ as
\begin{equation}
g_{(\mu\nu)}=g^{HI}_{(\mu\nu)}+\delta g_{(\mu\nu)},
\end{equation}
where $g^{HI}_{(\mu\nu)}$ denotes the homogeneous and isotropic solution of
$g_{(\mu\nu)}$ and $\delta g_{(\mu\nu)}$ are small quantities which
break the maximally symmetric solution with constant Riemannian curvature.
We shall simplify
our calculations by assuming that the density $\rho$ and the pressure $p$
only depend on the time $t$. Moreover, we assume that the mass parameter
$\mu \approx 0$ and we neglect any effects due to the antisymmetric source
tensor $B^{[\mu\nu]}$.

It is assumed that a solution can be found by a separation of variables
\begin{equation}
\label{separationeq}
\alpha(r,t)=h(r)R^2(t),\quad \beta(r,t)=r^2S^2(t).
\end{equation}

From the field equations, we get
\begin{equation}
\label{RSequation}
\frac{{\dot R}}{R}-\frac{{\dot S}}{S}=\frac{1}{2}Zr,
\end{equation}
where ${\dot R}=\partial R/\partial t$ and $Z$ is given by
\begin{equation}
Z=\frac{\dot\beta'f^2}{\beta^3}-\frac{5\dot\beta\beta' f^2}{2\beta^4}
-\frac{\dot\alpha\beta' f^2}{2\alpha\beta^3}+\frac{2\dot\beta ff'}{\beta^3}
-\frac{f\dot f'}{\beta^2}-\frac{3f'\dot f}{2\beta^2}
$$ $$
+\frac{\dot\alpha ff'}{2\alpha\beta^2}+\frac{2\beta'f\dot f}{\beta^3}.
\end{equation}
Let us assume that $Z\approx 0$, then from (\ref{RSequation}) we find
that $R\approx S$ and the metric line-element takes the form
\begin{equation} \label{FRWmetric}
ds^2=dt^2-R^2(t)\biggl[h(r)dr^2+r^2(d\theta^2+\sin^2\theta d\phi^2)\biggr].
\end{equation} From the conservation laws, we get \begin{equation} {\dot
p}=\frac{1}{R^3}\frac{\partial}{\partial t}[R^3(\rho+p)]. \end{equation}

If we assume that $\beta\gg f$, then the equations of motion
become~\cite{Moffat4,Moffat5}
\begin{equation}
\label{eqn1}
2b(r)+{\ddot{R}}(t)R(t)+2{\dot{R}}^2(t)-R^2(t)W(r,t)=4\pi
R^2(t)[\rho(t)-p(t)],
\end{equation}
\begin{equation}
\label{eqn2} -{\ddot
R}(t)R(t)+\frac{1}{3}R^2(t)Y(t)=\frac{4\pi}{3}R^2(t) [\rho(t)+3p(t)],
\end{equation}
where
\[ 2b(r)={h^\prime(r)\over r h^2(r)}. \]
The functions
$W$ and $Y$ are given by
\begin{equation}
\label{Wexpression}
W=\frac{\alpha'\beta'f^2}{2\alpha^2\beta^3}
-\frac{\beta^{\prime\prime}f^2}{\alpha\beta^3}
+\frac{\dot\alpha\dot\beta f^2}{2\alpha\beta^3} +\frac{5\beta'^2f^2}
{2\alpha\beta^4}-\frac{\dot\alpha f\dot f}{2\alpha\beta^2}
$$ $$
-\frac{\alpha'ff'}{2\alpha^2\beta^2}-\frac{ff^{\prime\prime}}{\alpha\beta^2}
-\frac{4ff'\beta'}{\alpha\beta^3}+\frac{3f'^2}{2\alpha\beta^2},
\end{equation}
\begin{equation}
\label{Yexpression} Y=\frac{\ddot\beta
f^2}{\beta^3}-\frac{5\dot\beta^2f^2}{2\beta^4} -\frac{3\dot
f^2}{2\beta^2}+\frac{4\dot\beta f\dot f}{\beta^3} -\frac{f\ddot
f}{\beta^2}.
\end{equation}
Within our approximation scheme, $W$ and $Y$ can be expressed in the form
\begin{equation}
\label{moreW}
W=\frac{h'f^2}{h^2r^5R^6}-\frac{2f^2}{hr^6R^6}+\frac{2{\dot
R}^2f^2}{r^4R^6}+\frac{10f^2}{hr^6R^6}-\frac{{\dot R}f{\dot f}}{r^4R^5}
-\frac{h'ff'}{2h^2r^4R^6}
$$ $$
-\frac{ff''}{h4^4R^6}-\frac{8ff'}{hr^5R^6}
+\frac{3f^{'2}}{2hr^4R^6},
\end{equation}
\begin{equation}
\label{moreY} Y=\frac{2({\dot R}^2+R{\ddot
R})f^2}{r^4R^6}-\frac{10{\dot R}^2f^2}{r^4R^6}-\frac{3{\dot
f}^2}{2r^4R^4}+\frac{8{\dot R}f{\dot f}}{r^4R^5} -\frac{f{\ddot
f}}{r^4R^4}.
\end{equation}

Eliminating ${\ddot R}$ by adding (\ref{eqn1}) and
(\ref{eqn2}), we get
\begin{equation}
\label{Rvelocityeq}
{\dot{R}}^2+b={8\pi\over 3}\rho R^2+QR^2,
\end{equation}
where
\begin{equation}
Q=\frac{1}{2}W-\frac{1}{6}Y.
\end{equation}
From (\ref{eqn2}) we obtain
\begin{equation}
\label{acceleration}
{\ddot R}=-\frac{4\pi}{3}R(\rho+3p)+\frac{1}{3}RY.
\end{equation}

We can write Eq.(\ref{Rvelocityeq}) as
\begin{equation}
H^2+\frac{b}{R^2}=\Omega H^2,
\end{equation}
where $H={\dot R}/R$,
\begin{equation}
\Omega=\Omega_M+\Omega_Q,
\end{equation}
and
\begin{equation}
\Omega_M=\frac{8\pi\rho}{3H^2},\quad \Omega_Q=\frac{Q}{H^2}.
\end{equation}
If $b=0$, then we get $\Omega=1$ and
\begin{equation}
\label{FlatFried}
H^2=\frac{8\pi}{3}\rho+Q.
\end{equation}
The line element now takes the approximate form of a flat, homogeneous and
isotropic FRW universe
\begin{equation}
ds^2=dt^2-R^2(t)[dr^2+r^2(d\theta^2+\sin^2\theta d\phi^2)].
\end{equation}

\section{Accelerating Expansion of the Universe}

It follows from (\ref{acceleration}) that ${\ddot R} > 0$ when $Y >
4\pi(\rho+3p)$. If we assume that there is a
solution for $Q$ and $Y$, such that they are small and constant in the
early universe, then we will retain the good
agreement of GR with the BBN era with
$\rho_{\rm rad}\propto 1/R^4$ and $\rho_M\propto 1/R^3$. As the universe
expands beyond the BBN era at the temperatures, $T\sim 1$ MeV-$60$ kev,
then $Q$ begins to increase and reaches a slowly varying value with
$\Omega^0_Q\sim 0.7$ and $\Omega^0_M\sim 0.3$, where $\Omega^0_M$ and
$\Omega^0_Q$ denote the present values of $\Omega_M$ and $\Omega_Q$,
respectively. These values can fit the combined supernovae, cluster and CMB
data~\cite{Perlmutter}.

We observe from (\ref{moreW}) and (\ref{moreY})
that the dependence of $Q$ and $Y$ as the universe expands is a function of
the behaviour of $R$ and $f$ and their derivatives. If
$f$ grows sufficiently with $R$ as $t$ increases, then $Q$ and $Y$ can
dominate the matter contribution $\rho_M$ as the universe evolves towards
the current epoch. A detailed solution of the field equations is required
to determine the dynamical behaviour of $R$, $f$, $Q$ and $Y$. However, we
can obtain some knowledge of the qualitative behaviour of $Y$ and $Q$ by
assuming that ${\dot f}\sim f'\sim 0$ and $h=1$. Then, from (\ref{moreW})
and (\ref{moreY}) we obtain
\begin{equation}
Y\approx \frac{2f^2}{r^4R^6}(R{\ddot R}-4{\dot R}^2),
\end{equation}
and
\begin{equation}
Q\approx \frac{f^2}{3r^4R^6}\biggl(\frac{12}{r^2}+7{\dot R}^2-R{\ddot
R}\biggr).
\end{equation}
If, as the universe expands, the behaviour of $f$ is $f\sim
R^a$ with $a\ge 2$, then we can satisfy $Y >0$, $Q>0$, ${\ddot R} >0$ and
${\dot R}$ will increase as we approach the present epoch corresponding to
an accelerating expansion.

We can explain the evolution of Hubble expansion acceleration
within NGT, without violating the positive energy conditions. Both $\rho$
and $p$ remain positive throughout the evolution of the universe. There is
no need for a dark energy and a cosmological constant. Thus, we avoid
having to explain the unnatural and mysterious ``coincidence'' of matter
and dark energy domination. The $Q$ contribution to the
expansion of the universe increases at a slow rate up to a constant value
today with $Y >4\pi\rho$ ($p\approx 0$), and can then decrease to zero as
the universe continues to expand, avoiding an eternally accelerating
universe. During this evolution, the cosmological constant
$\Lambda=0$. It is then possible to avoid the existence of a
cosmological horizon and the problems it produces for quantum field
theory and string theory~\cite{Fischler}.

\section{NGT and Dark Matter}

Galaxy dynamics observations continue to pose a problem for gravitational
theories and cosmology. The data for spiral galaxies
are in sharp contradiction with Newtonian dynamics, for virtually all
spiral galaxies have rotational velocity curves which tend towards a
constant value. The standard assumption is that dark matter exists in
massive, almost spherical halos surrounding galaxies. The standard
hypothesis is that about 90\% of the mass is in the form of dark matter and
dark energy and this explains the flat rotational velocity curves of
galaxies. This explanation is not economical, for it requires three or more
parameters to describe different kinds of galactic systems and no
satisfactory model of galactic halos exists~\cite{Sellwood}.

A possible explanation of the galactic rotational velocity curves problem
has been obtained in NGT~\cite{Sokolov}. A derivation of the motion of test
particles yields the total radial acceleration experienced by a test
particle in a static spherically symmetric gravitational field for $r\ge
0.2$ kpc, due to a point source (we reinsert $G$ and $c$)~\cite{Legare}:
\begin{equation}
a(r)=-\frac{GM}{r^2}+\frac{\lambda
Cc^2}{r^2_0}\frac{\exp(-r/r_0)}{r^2}\biggl(1+\frac{r}{r_0}\biggr),
\end{equation}
where $\lambda$, $C$ and $r_0=1/\mu$ are constants which
remain to be fixed.

We choose $C\propto \sqrt{M}$ and fix $\lambda$ to give
\begin{equation}
a(r)=-\frac{G_{\infty}M}{r^2}+G_0\sqrt{M}\frac{\exp(-r/r_0)}{r^2}
\biggl(1+\frac{r}{r_0}\biggr),
\end{equation}
where $G_{\infty}$ is defined to be the gravitational constant at infinity
\begin{equation}
G_{\infty}=G_0\biggl(1+\sqrt{\frac{M_0}{M}}\biggr)
\end{equation}
and $G_0$ is Newton's gravitational constant.

These formulas were applied to explain the flatness of rotation curves of
galaxies, as well as the Tully-Fisher law~\cite{Tully}, $L\sim v^4$, where
$v$ is the rotational velocity of a galaxy and $L$ is the luminosity. A
derivation of $v$ gives
\begin{equation}
v^2=\frac{3G_0L}{r}\biggl\{1+\sqrt{\frac{L_0}{L}}[1-\exp(r/r_0)
\biggl(1+\frac{r}{r_0}\biggr)]\biggr\},
\end{equation}
where  $L_0=3M_0$. For distances less than $0.5-4$ kpc, the standard Newtonian
law of gravity will apply. For $r_0=25$ kpc and $L_0=250\times
10^{10}\,L_{\odot}$, an excellent fit to spiral galaxies was
found~\cite{Sokolov,Sokolov2}. Moreover, a good fit to the Tully-Fisher law
was also obtained.

Consider now the giant spiral galaxy M31 and our Galaxy in the local
group. The center of M31 is approaching the center of the Galaxy at a
velocity $\sim 119$ km/sec.
The total mass of the local group should be very large, namely, the
mass-to-light ratio should be $\sim 100 \biggl(M/L_{\odot}\biggr)$. This
big ratio is normally explained using the dark matter hypothesis. The
distance between M31 and the Galaxy is $\sim 700$ kpc, so the additional
exponential force is vanishingly small, but what is left is the
renormalized gravitational constant. Thus, the gravitational acceleration
becomes
\begin{equation} a(r)=-\frac{G_0M^*}{r^2},
\end{equation}
where
$M^*\sim 17M$ and from the observed mass-to-light ratio:
\begin{equation}
\frac{M^*}{L}\sim 100\biggl(\frac{M}{L}\biggr)_{\odot},
\end{equation} we
predict
\begin{equation} \frac{M}{L}\sim
6\biggl(\frac{M}{L}\biggr)_{\odot}.
\end{equation}
This agrees with the estimated ratio for luminous matter without using the
dark matter assumption.

Gravitational lensing effects can also be accounted for in NGT. We find
for the angle of deflection $\Delta\phi$, obtained in the post-Newtonian
approximation
\begin{equation}
\Delta\phi=\frac{4G_0\biggl(1+\sqrt{M_0/M}\biggr)}{c^2R},
\end{equation}
where $M$ is the mass of the galaxy and $R$ is the distance between the
galaxy center and the deflected ray. This prediction is close to the one
obtained from the dark matter hypothesis.

If we calculate the acceleration expected in our solar system, we obtain
\begin{equation}
\delta a=\frac{a-a_{\rm NGT}}{a_{\rm Newton}}\approx
\frac{1}{2}\frac{\sqrt{M_0}}{M}\biggl(\frac{r}{r_0}\biggr)^2.
\end{equation}
For solar and terrestrial experiments, we find for $r_0\sim 25$ kpc,
$\delta a < 10^{-13}$, which is too small a deviation from Newton's law to
be detected with current experiments.

We must still acount for the estimated value of $\Omega_M\sim 0.33\pm
0.053$~\cite{Turner}. Measurements of total baryon density give
$\Omega_Bh^2=0.020\pm 0.001$~\cite{Burles}. For $h\sim
0.7$ ,we get $\Omega_B\sim0.04$. The usual hypothesis states that cold
dark matter particles contribute $\Omega_{\rm CDM}\sim 0.3$. We do expect
that there is some dark matter in the universe in the form of dark baryons
and neutrinos with non-vanishing mass ( $\leq10\%$ ). It remains to be seen
whether an alternative gravity theory such as NGT can provide an
explanation for the discrepancy between visible baryon matter and dark
matter. At the era of structure formation, the contribution of $\Omega_Q$
could be of order $\Omega_Q\sim 0.3$, growing to its present day value of
$\Omega^0_Q\sim 0.95$ so that $\Omega^0\sim 1$. NGT would then be
required to explain the formation of galaxy structure without CDM. These are issues
that require further investigation.

\vskip 0.2 true in
{\bf Acknowledgments} \vskip 0.2 true in

This work was supported by the Natural Sciences and Engineering Research
Council of Canada.
\vskip 0.5 true in


\begin{thebibliography}{100}
 
\bibitem{Perlmutter} S. Perlmutter et al. Ap. J. {\bf 483},
565 (1997), astro-ph/9608192; A. G. Riess, et al. Astron. J. {\bf 116},
1009 (1998), astro-ph/9805201; P. M. Garnavich, et al. Ap. J. {\bf 509}, 74
(1998), astro-ph/9806396; S. Perlmutter et al. Ap. J. {\bf 517}, 565
(1999), astro-ph/9812133; A. G. Riess, et al. to be published in Ap. J.,
astro-ph/0104455.
\bibitem{Caldwell} R. R. Caldwell, R. Dave, and P. J. Steinhardt, Phys.
Rev. Lett. {\bf 80}, 1582 (1998), astro-ph/9708069.
\bibitem{Woodard} M. Brisudova and R. P. Woodard, gr-qc/0105072 v2.
\bibitem{Sellwood} J. A. Sellwood and A. Kosowsky, astro-ph/0009074.
\bibitem{Fischler} W. Fischler, A. Kashani-Poor. R. McNees and S. Paban,
hep-th/0104181; S. Hellerman, N. Kaloper, and L. Susskind,
JHEP 0106 (2001) 003, hep-th/0104180.
\bibitem{Cline} J. W. Moffat,
hep-th/0105017 v2; J. M. Cline, hep-th/0105251 v2; E. Haylo,
hep-ph/0105216; C. Kolda and W. Lahneman, hep-th/0105300; C. Deffayet, G.
Dvali and G. Gabadadze, hep-th/0105068; J. Ellis, N. E. Mavromatos, and D.
V. Nanapolous, hep-th/0105206, Xiao-Gang He, hep-th/0105005 v2.
\bibitem{Carroll} S. M. Carroll, astro-ph/0107571; S. M. Carroll, and M.
Kaplinghat, astro-ph/0108002.
\bibitem{Weinberg} S. Weinberg, {\it
Gravitation and Cosmology: Principles and Applications of the General
Theory of Relativity}, John Wiley and Sons, New York, 1972.
\bibitem{Binetruy} P. Binetruy, C. Deffayet and D. Langlois, Nucl. Phys.
{\bf B565}, 269 (2000), hep-th/9905012; D. J. H. Chung and K. Freese,
Phys.Rev. {\bf D61} {\bf 61} (2000) 023511, hepph/9910235; C. Csaki, M.
Graesser, C. Kolda and L. Terning, Phys. Rev. Lett. {\bf B462}, 34 (1999),
hep-ph/9906513, K. Maeda, astro-ph/0012313; G. Huey and J. E. Lidsey,
astro-ph/0104006.
\bibitem{Moffat} J. W. Moffat, Phys. Letts. {\bf B 335},
447 (1995); J. W. Moffat, J. Math. Phys. {\bf 36}, 3722 (1995); Erratum, J.
Math. Phys. {\bf 36}, 7128 (1995).
\bibitem{Moffat2} J. W. Moffat, J. Math.
Phys. {\bf 36}, 5897 (1995).
\bibitem{Moffat3} J. W. Moffat, gr-qc/9605016
v3.
\bibitem{Moffat4} J. W. Moffat, astro-ph/9510024 v2.
\bibitem{Moffat5}
J. W. Moffat, astro-ph/9704300.
\bibitem{Damour} T. Damour, S. Deser, and
J. McCarthy, Phys. Rev. {\bf D47}, 1541 (1993).
\bibitem{Clayton} M. A.
Clayton, Class. Quant. Grav. {\bf 13}, 2851 (1996), gr-qc/9603062; J. Math.
Phys. {\bf37}, 395 (1996), gr-qc/9505005;  Int. J.
Mod. Phys. {\bf A12}, 2437 (1997), gr-qc/9509028; {\it Massive Nonsymmetric
Gravitational Theory: A Hamiltonian Approach}, Ph.D. thesis, University of
Toronto, 1996.
\bibitem{Sokolov} J. W. Moffat and I.
Yu. Sokolov, Phys. Lett. {\bf B378}, 59 (1996), astro-ph/9509143 v3.
\bibitem{Sokolov2} Fits to more spiral galaxies and clusters have been
obtained that agree well with the data, private communication, I. Yu.
Sokolov.
\bibitem{Legare} J. L\'egar\'e and J. W. Moffat, Gen. Rel. and
Grav. {\bf 27}, 761 (1995); gr-qc/9412009; gr-qc/9509035.
\bibitem{Tully}
R. B. Tully and J. R. Fisher, Astr. Ap. {\bf 54}, 661 (1977).
\bibitem{Turner} M. S. Turner, Physica Scripta {\bf T85}, 210, 2000.
\bibitem{Burles} S. Burles, K. M. Nollett, and M. S. Turner, ApJ, {\bf
552}, L1; C. Pryke, et al., astro-ph/0104490; C. B. Netterfield et al.,
astro-ph/0104460.
\end{thebibliography}
\end{document}